\documentclass[twocolumn,amsmath,amssymb,aps,pra,superscriptaddress,tightenlines,10pt,conference]{revtex4}
\usepackage{lipsum}
\usepackage{amsmath}
\usepackage{amsfonts}
\usepackage{graphicx}
\usepackage{epsfig}
\usepackage{color}
\usepackage{bm}
\usepackage{color}
\usepackage{mathtools}
\usepackage[colorlinks,citecolor=blue,linkcolor=magenta]{hyperref}
\usepackage[normalem]{ulem}
\begin{document}
\title {Interplay of quantum phase transition and flat band in hybrid lattices }

\author{Gui-Lei Zhu}
\affiliation{School of Physics, Huazhong University of Science and Technology, Wuhan 430074, China}

\author{Hamidreza Ramezani}
\affiliation{Department of Physics and Astronomy, University of Texas Rio Grande Valley, Edinburg, Texas 78539, USA}

\author{Clive Emary}
\affiliation{Joint Quantum Centre (JQC) Durham-Newcastle, School of Mathematics, Statistics and Physics, Newcastle University,
Newcastle-Upon-Tyne, NE1 7RU, United Kingdom}

\author{Jin-Hua Gao}
\affiliation{School of Physics, Huazhong University of Science and Technology, Wuhan 430074, China}

\author{Ying Wu}
\affiliation{School of Physics, Huazhong University of Science and Technology, Wuhan 430074, China}

\author{Xin-You L\"{u}}
\email{xinyoulu@hust.edu.cn}
\affiliation{School of Physics, Huazhong University of Science and Technology, Wuhan 430074, China}
\date{\today}
\begin{abstract}
{We establish a connection between quantum phase transitions (QPTs) and energy band theory in an {\it extended} Dicke-Hubbard lattice, where the periodical critical curves modulated by wave number $k$ leads to rich equilibrium dynamics. Interestingly, the chiral-symmetry-protected flat band and the localization that it engenders, exclusively occurs in the normal phase, and disappears in the superradiant phase. This originates from that QPT breaks up the on-site resonance condition and off-site chiral symmetry of system simultaneously, which prohibits the destructive interference for obtaining a flat band. Our work offers an approach to identify different phases of lattice via detecting the flat bands or simply the related localizations in a single cell, and in turn, to control the appearance of flat bands by QPT. }
\end{abstract}
\maketitle
\section{introduction}
The quantum phase transition (QPT), driven by quantum fluctuations in many-body systems, is one of the most fundamental and significant concepts in physics, since it can offer the important resources for quantum metrology\,\cite{Macieszczak2016,Farbe,Lorenzo2017} and quantum sensing\,\cite{Goldstein2010,Raghunandan2018,Zhang2018}. For example,  the generation of many-body entanglement through QPT enables precision metrology to reach the Heisenberg limit\,\cite{You2017,Huang2018}. To apply QPT theory into modern quantum technologies, a necessary task is to determine in which phase the system is. Traditionally, one needs to detect the order parameters based on the ground states of system\,\cite{Sachdev}, which is experimentally challenging in lattice systems, where the complexity of ground states increases exponentially with the size of the system. On top of that, the detecting precision is normally restricted by inevitable experimental uncertainties and fabrication errors, e.g., small perturbations. Hence it is highly demanded to develop a robust method to precisely identify different phases of a quantum lattice system in the absence of ground state detection.

In past decade and on a different subject, flat bands (corresponding to a zero group velocity and an infinite effective mass) have been extensively studied in various condensed-matter contexts \cite{Lieb1989,Vidal1998,Bergman2008,Ge2015,Qiu2016,Ozawa2017,Hamid2017}, on account of its potential applications in realizing fractional quantum Hall effect in the absence of Landau levels~\cite{Tang2011,Sun2011,Neupert2011,Sheng2011,Regnault2011}, engineering strong nonlinear correlations\,\cite{Biondi2015} and diffraction-free transmission of light\,\cite{silva2014}. Recently, the flat bands have been observed in experiments with exciton-polariton condensates\,\cite{Jacqmin2014,Baboux2016}, photonic lattices\,\cite{Li2008,Vicencio2015,Mukheriee2015,alex19} and cold atom lattices\,\cite{Taie2015}. With an excellent and unique property,  such a flat band is robust against perturbations of system parameters, which opens up the opportunity for detecting phases of matter with strong robustness.
 
\begin{figure}
\centerline{\includegraphics[width=8.4cm]{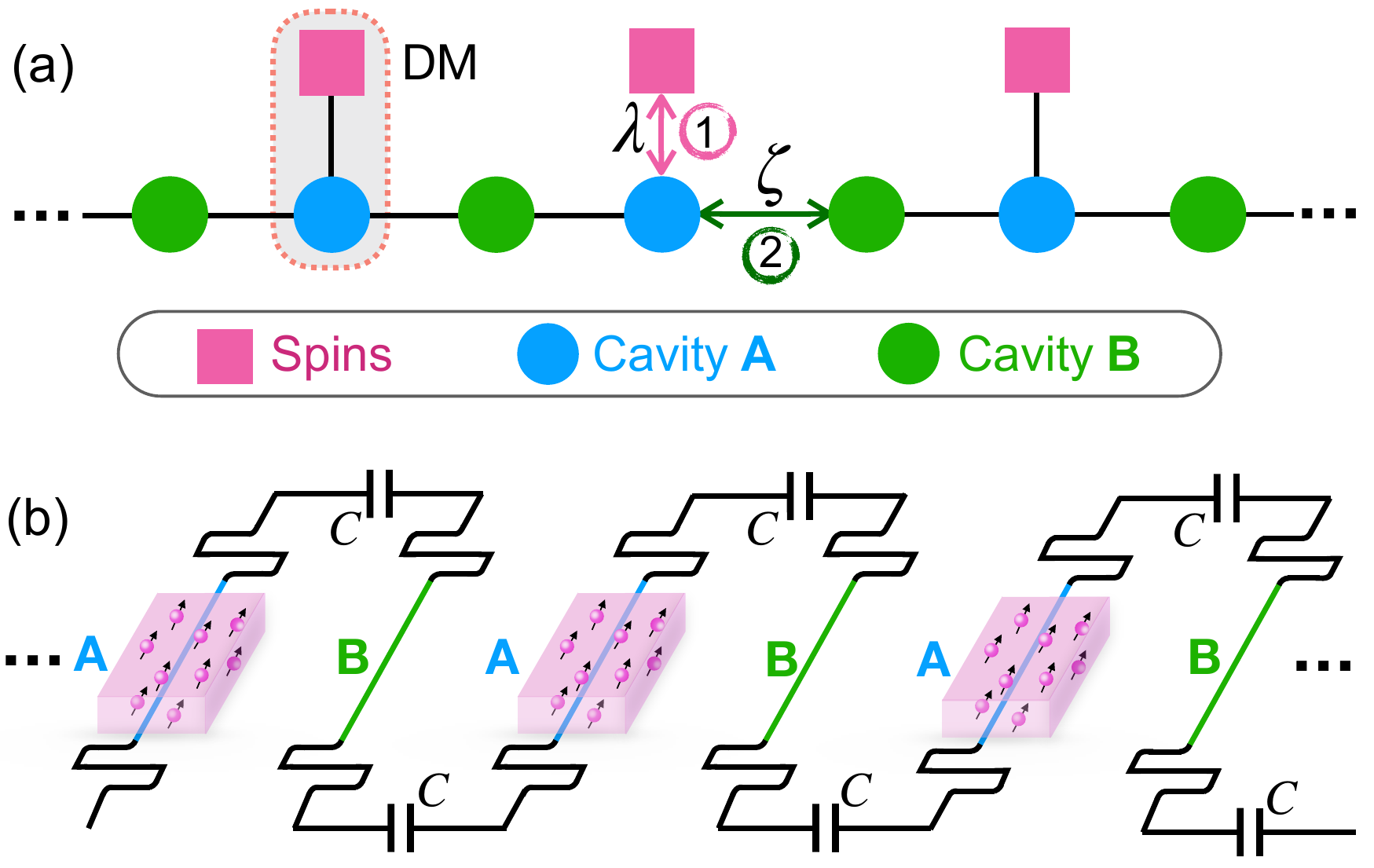}}
\caption{(a) Schematic illustration of one dimensional (1D) {extended} Dicke-Hubbard lattice, where the unit cell consists of a Dicke model (spin ensemble collectively interacting with cavity $A$) coupled to an atomless cavity $B$.  Cavity $A$ interacts with the spin ensemble and cavity $B$ with coupling strengths $\lambda$ and $\zeta$, respectively, forming two symmetric channels of the lattice,  labelled by 1 and 2. (b) The implementation of {extended} Dicke-Hubbard lattice in a hybrid superconducting circuit with capacitance coupling labelled by $C$ and the ensemble of spins in a diamond crystal.} 
\label{fig1}
\end{figure}

Here we established the connection between flat band and the QPT from normal phase to the superradiant phase in an {\it extended} Dicke-Hubbard lattice, i.e., a series of Dicke models\,\cite{Dicke1954} coupled together through a set of atomless cavities. Experimentally, this extended model can be implemented in hybrid superconducting circuits\,\cite{Bensky2011,Houck2012,Underwood2012,xinyou2013,Ze-Liang,Zou2014,Li2015}, in which two-level ensemble (e.g., NV center spins) is doped in every other cavity [see Fig.\,\ref{fig1}(b)]. The superradiant QPT (a second-order phase transition) was proposed in the single Dicke model, and occurs when increasing the atom-field coupling through a critical point\cite{Hepp1973,Wang1973,Emary2003,Lambert2004,Baumann2010,Nagy2010,Nataf2010,Klinder2015,Hanpu2016,Liu2018,Xin-You2018,Hanpu2019,Guilei2019}, which is associated with a spontaneously $\mathbb{Z}_{2}$ symmerty breaking. Extending to the periodic lattice, however, we here find this critical point is replaced by the critical curves periodically modulated by wave number $k$. The periodical boundaries of normal and superradiant phases intersect at some certain values of $k$. This predicts, in the lattice systems, a critical region between the normal and superradiant phases, where the first-order phase transition and unstable phases alternatively appear in the different range of $k$. 

The above connection allows us to identify different phases of system via experimentally detectable energy bands (or the occupation localization in a single cell), and in turn, to control the occurrence of flat bands using QPT. Specifically, in the normal phase, a chiral-symmetry-protected flat band appears in the spectrum of system, which features asymmetric band structures originated from the counter-rotating interactions. This flat band disappears once the system enters into the superradiant phase.  Physically, the superradiant QPT makes both the excitation energy and an additional potential of spin ensemble interaction-dependent, which breaks up the on-site resonant condition and off-site chiral symmetry of lattice, respectively.  Either of them can destroy the destructive interference of lattice, and finally lead to the disappearance of flat band.  

\section{Superradiant phase transition in 1D lattice system}\label{sec2}
As illustrated in Fig.\,\ref{fig1}, we consider a 1D {extended} Dicke-Hubbard lattice implemented by a hybrid superconducting circuit, with the Hamiltonian, 
\begin{align}
{\mathcal H}=&\sum_{n_A}{\mathcal H}_{n_A}^{\rm Dicke}
+\sum_{n_B} {\mathcal H}_{n_B}^{\rm Cavity}
+{\mathcal H}_{\rm int},
\end{align}
where the subscript $n_A$ ($n_B$) denotes the lattice site of cavity $A$ ($B$). The Dicke Hamiltonian is given by ${\mathcal H}_{n_A}^{\rm Dicke}= \omega_{A}a_{n_A}^{\dagger}a_{n_A}+\Omega J_{n_A}^{z}+\frac{\lambda}{\sqrt{N}}(a_{n_A}^{\dagger}+a_{n_A})(J_{n_A}^{+}+J_{n_A})$, where $a_{n_A}$ is annihilation operator of cavity $A$ mode, and $J_{n_A}^{z}=(1/2)\sum_{N}\sigma_{n_A}^{z}$, $J_{n_A}^{\pm}=\sum_{N}\sigma_{n_A}^{\pm}$ are the collective operators of $N$ spins. The Hamiltonian ${\mathcal H}_{n_B}^{\rm Cavity} = \omega_{B} a_{n_B}^{\dagger}a_{n_B}$ describes the bare cavity $B$. The nearest-neighbor cavities are coupled via an $x$-$x$ interaction ${\mathcal H}_{\rm int}=-\zeta\sum_{\langle n_A,n_B\rangle}(a_{n_A}^{\dagger}+a_{n_A})(a_{n_B}^{\dagger}+a_{n_B})$. 

To explore the phase transition in this  extended Dicke-Hubbard lattice, we extend and modify the method in Ref.~\cite{Emary2003} to the lattice systems. With the Holstein-Primakoff transformation, i.e., $J_{n_A}^{+}=b_{n_A}^{\dag}\sqrt{N-b_{n_A}^{\dag}b_{n_A}},\,J_{n_A}^{-}=\sqrt{N-b_{n_A}^{\dag}b_{n_A}}\,b_{n_A},$ and $J_{n_A}^{z}=b_{n_A}^{\dag}b_{n_A}-N/2$, here we introduced the bosonic operators $b_{n_A}$ obeying $[b_{n_A},b_{n_A}^{\dag}]=1$.  In the thermodynamic limit $N \to \infty$, the system Hamiltonian in $k$ space can be obtained by the Fourier transformation $O_n=\frac{1}{\sqrt{\mathcal N}}\sum_{k}e^{ik\cdot n}O_k$ ($O_{n}$ is an arbitrary annihilation operator and ${\mathcal N}$ is the number of unit cells). In the normal phase, Hamiltonian is given by 
$\mathcal{H}_{\rm nor}(k)={1}/{2}\sum_{k} \psi^{\dag}_{\rm nor}{{\mathcal M_{\rm nor}}(k)}\psi_{\rm nor},$
where $\psi_{\rm nor}=[a_{kA},a_{kB},b_{kA},a_{-kA},a_{-kB},b_{-kA}]^{\rm T}$ and the superscript ${\rm T}$ denote a transpose operation. The coefficient is collected into the matrix
\begin{center}\begin{equation}
\mathcal{M}_{\rm nor}(k)=\left(\begin{array}{cccccc}
\omega_A&f&\lambda&0&f&\lambda \\
f^*&\omega_B&0&f^*&0&0 \\
\lambda&0&\Omega&\lambda&0&0\\
0&f&\lambda&\omega_A&f&\lambda\\
f^*&0&0&f^*&\omega_B&0\\
\lambda&0&0&\lambda&0&\Omega
\end{array}\right)
,
\label{matrix1}
\end{equation}\end{center}
where $f=-\zeta[1+{\rm exp}(ik)]$. Here we have taken the lattice constant to be identical and a periodic boundary condition. The matrix $\mathcal{M}_{\rm nor}(k)$ can be divided into the on-site part $\mathcal{M}^{\rm on}_{\rm nor}(k)=\rm {diag}\{\omega_A,\omega_B,\Omega,\omega_A,\omega_B,\Omega\}$ and off-site interaction part $\mathcal{M}^{\rm int}_{\rm nor}(k)=\mathcal{M}_{\rm nor}(k)-\mathcal{M}^{\rm on}_{\rm nor}(k)$. Especially, the interaction matrix $\mathcal{M}^{\rm int}_{\rm nor}(k)$ satisfies the chiral symmetry with ${\mathcal C^{\dag}}\mathcal{M}^{\rm int}_{\rm nor}(k){\mathcal C}=-\mathcal{M}^{\rm int}_{\rm nor}(k)$, where ${\mathcal C}=\rm {diag}\{-1,1,1,-1,1,1\}$, and this symmetry is exact in the thermodynamic limit.  Hamiltonian ${\mathcal H}_{\rm nor}(k)$ is bilinear in terms of bosonic operators, which can be diagonalized analytically. To analytically diagonalize the Hamiltonian $\mathcal{H}_{\rm nor}(k)$, firstly, we introduce an ancillary matrix ${\mathcal D}_{\rm nor}=\tau_z \mathcal{M}_{\rm nor}$, where $\tau_z={\rm diag}\{1,1,1,-1,-1,-1\}$. Secondly, we apply the transformation matrix $\mathcal{T}$ into Hamiltonian ${\mathcal H}_{\rm nor}(k)$, where $\mathcal{T}$ simultaneously satisfies
\begin{center}\begin{equation}
	{\mathcal T^{-1} \mathcal D_{\rm nor}\mathcal T}=\left(\begin{array}{cc}
	E_{\rm nor}(k)&0\\
	0&-E_{\rm nor}(-k)
	\end{array}\right)
	\end{equation}\end{center}
and ${\mathcal T^{\dag} }\tau_z{\mathcal T}=\tau_z$. Lastly, based on the above method and considering $\Omega=\omega_A=\omega_B=\omega$,  the energy spectra of $\mathcal{H}_{\rm nor}(k)$ are
\begin{equation}
E_{\rm nor}(k)=\begin{cases}
{\color{black}\sqrt{\omega^2-2\omega\sqrt{2\zeta^2({1+\rm cos}\, k)+\lambda^2}}};\\
\omega;\\
{\color{black}\sqrt{\omega^2+2\omega\sqrt{2\zeta^2(1+{\rm cos}\, k)+\lambda^2}}}.
\end{cases}\label{eigenval1}
\end{equation}
Note that both the on-site resonance condition and off-site chiral symmetry ultimately lead to a robust flat band located at $E_{\rm nor}(k)=\omega$  in our model (detailed discussion is shown below). Here the  lowest excitation energy $E^{l}_{\rm nor}(\lambda,k)=\sqrt{\omega^2-2\omega\sqrt{2\zeta^2(1+{\rm cos}\,k)+\lambda^2}}$ indicates that our model is well defined when $\left|{\zeta}/{\omega}\right|<1/4$.
Beyond this regime, the Hamiltonian does not possess normalizable eigenfunctions and has no obvious physical meaning for all values of other system parameters\,\cite{Emary2002}. 

\begin{figure*}
	\centerline{\includegraphics[width=17.0cm]{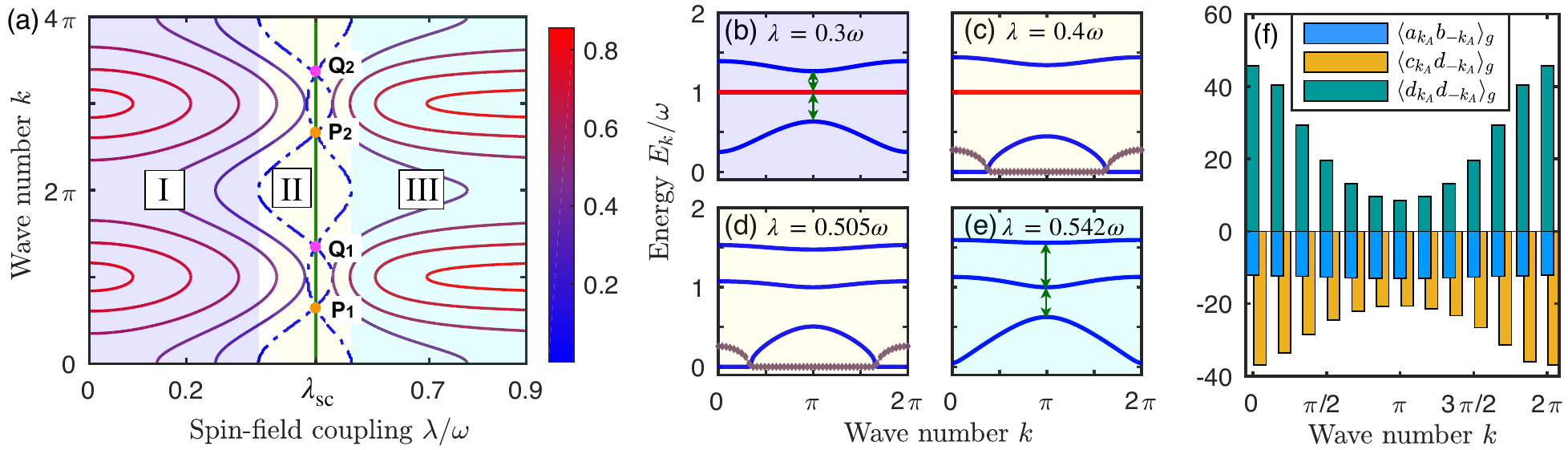}}
	\caption{(a) Contour plots of the lowest excitation energies $E_{\rm nor}^{l}(\lambda,k)$ and  $E_{\rm sup}^{l}(\lambda,k)$. The dot-dashed curves are related to $E_{\rm nor}^{l}(k,\lambda)=0$ and $E_{\rm sup}^{l}(k,\lambda)=0$, which are the boundaries of normal phase (region I)  and superradiant phase (region III). They intersect at points $(\lambda_{\rm sc}, P_{n})$ and $(\lambda_{\rm sc}, Q_{n})$, $(n=0,\pm1,\pm2,...)$ with $P_{n}=2n\pi-4\pi/3$, $Q_{n}=2n\pi-2\pi/3$ and $\lambda_{\rm sc}$ being the critical point of a single cell. This leads to a critical region labelled by region II in that for some certain $k$, i.e., $k\in[Q_{n},P_{n+1}]$, the lowest excitation energies become merely imaginary. (b-e) Band structures of 1D  extended Dicke-Hubbard lattice. The solid (diamond) curves show the real (imaginary) part of the energy. (f) The expected values of three pairing operators between $k$ and $-k$ spaces in the ground state. The considered system parameters are $\Omega=\omega_A=\omega_B=\omega=1$ and $\zeta=0.18\omega$.}  
	\label{fig2}
\end{figure*}

The system undergoes a superradiant phase transition when the lowest excitation energy $E^l_{\rm nor}(\lambda,k)= 0$ with increasing $\lambda$\,\cite{Emary2003}, and thus $E^l_{\rm nor}(\lambda,k)= 0$ provides the boundary of the normal phase. Beyond the regime of normal phase, the weak excitation approximation used in the derivation of $E^l_{\rm nor}(\lambda,k)$ is invalid, and $E^l_{\rm nor}(\lambda,k)$ becomes imaginary. Then we should make a macroscopic displacement on the bosonic modes, i.e., $a_{n_A}^{\dagger}\rightarrow c_{n_A}^{\dagger}+{\alpha}, \,\,\, b_{n_A}^{\dagger}\rightarrow d_{n_A}^{\dagger}-{\beta},\,\,\,a_{n_B}^{\dagger}\rightarrow c_{n_B}^{\dagger}+{\gamma}$ with
 \begin{align}
{\alpha}=&\pm\frac{\Omega}{2\mu\lambda}\sqrt{\frac{N}{4}(1-\mu^2)},\\ {\beta}=&\pm\sqrt{\frac{N}{2}(1-\mu)},\\ {\gamma}=&\frac{2\zeta}{\omega_B}\alpha,
\end{align} where \begin{align}\mu=\frac{\Omega(\omega_{A}-4\zeta^2/\omega_{B})}{4\lambda^2}.\end{align}
As the similar procedure used in normal phase, the displaced Hamiltonian $\mathcal{H}_{\rm sup}(k)={1}/{2}\sum_{k} \psi^{\dag}_{\rm sup}{{\mathcal M_{\rm sup}}(k)}\psi_{\rm sup}$ with  $\psi_{\rm sup}=[c_{k_A},c_{k_B},d_{k_A},c_{-k_A}^{\dag},c_{-k_B}^{\dag},d_{-k_A}^{\dag}]^{\rm T}$ and
\begin{center}\begin{equation}
	\mathcal{M}_{\rm sup}(k)=\left(\begin{array}{cccccc}
	\omega_A&f&\xi&0 &f& \xi \\
	f&\omega_B&0&f&0&0 \\
	\xi&0&\chi+2\eta&\xi&0&2\eta \\
	0&f&\xi&\omega_A&{\color{black}f}&\xi \\
	f&0&0&{\color{black}f}&\omega_B&0  \\
	\xi&0&2\eta&\xi&0&\chi+2\eta
	\end{array}\right), \label{matrix2}
	\end{equation}\end{center}
where $\chi={\Omega}(1+\mu)/{(2\mu)}, 
\xi=\lambda\mu\sqrt{{2}/{(1+\mu)}},
\eta={\Omega(1-\mu)(3+\mu)}/{[8\mu(1+\mu)]}.$
The analytical energy spectra $E_{\rm sup}(k)$ with complicated form are shown numerically, in which $E^l_{\rm sup}(\lambda,k)=0$ indicates the periodical boundary of superradiant phase.

To clearly show the phase transition in the present lattice system, in Fig.\,{\ref{fig2}}(a) we plot the contour of the real part of the lowest excitation energy versus spin-field coupling $\lambda$ and wave number $k$ in the normal phase and superradiant phase, respectively. Obviously, our model features lots of distinctive characters. First of all, both the normal and superradiant phases have the boundaries periodically modulated by wave number $k$. Physically, this demonstrates that, in the lattice system, the different spreading waves decided by various $k$ cause a periodic modulation on critical points of phase transition. Secondly,  the above two periodical boundaries intersect at the critical points of a single cell, i.e., $\lambda=\lambda_{\rm sc}=\sqrt{\Omega(\omega_A-4\zeta^2/\omega_B)}/2$, where the many-body effects of the periodical lattice system disappear. Correspondingly, $k$-coordinates of the crossing points satisfying $\cos k=-1/2$ are equivalent to $P_{n}=2n\pi-4\pi/3$ and $Q_{n}=2n\pi-2\pi/3$ ($n=0,\pm1,\pm2...$)[see Appendix], which divides the blank and overlap zones of the critical region. Physically, the blank zones corresponding to the lowest excitation energy being imaginary are the unstable phases. The overlap zones indicate that the system has an effective triple-well potential which allows first-order phase transitions~\cite{Hayn2011,Baksic2014,Soriente2018,Yang2020}. 
 Lastly, in a short summary, the present lattice system has three parameter regions, i.e., the normal phase, superradiant phase, and a critical region, where some certain spreading waves become dynamics unstable.

\section{Flat band associated QPT} 
\begin{figure}
	\centerline{\includegraphics[width=8.6cm]{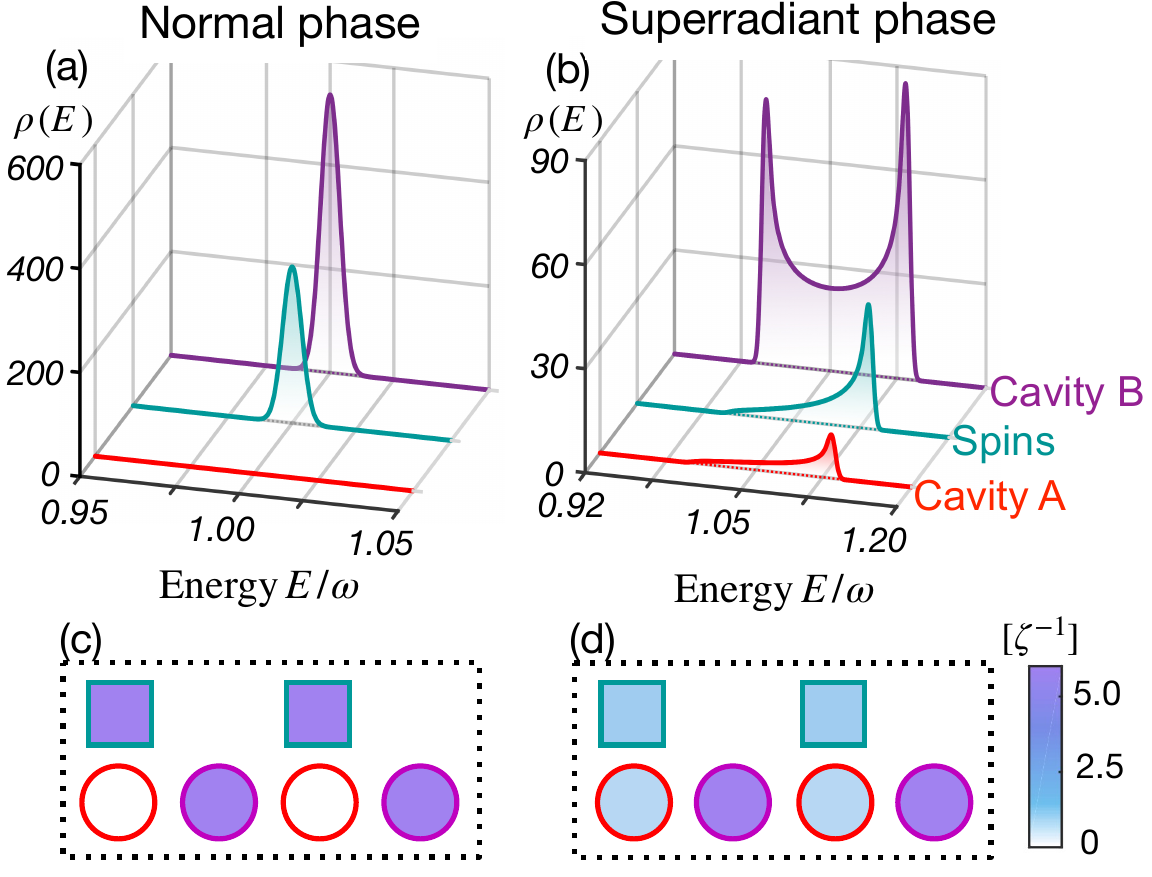}}
	\caption{(a,b) The local density of states (LDOS) of cavity modes $A$, $B$ and spins for the middle band of the 1D { extended} Dicke-Hubbard lattice. The corresponding mode profile in the real space shown in (c,d). The red (purple) circle and blue square denote cavity $A$ ($B$) and spins modes, respectively. (a,c) are in normal phase ($\lambda=0.3\omega$), and (b,d) are in the superradiant phase ($\lambda=0.542\omega$). Other parameters are same as Fig.\,\ref{fig2}. }
	\label{fig3}
\end{figure}

To show the interplay of the flat band and QPT, in Figs.\,\ref{fig2}(b-e), we plot the dispersion relation of 1D  extended Dicke-Hubbard lattice in the first Brillouin zone for different spin-field couplings. In the normal phase, a flat band locates at $E_{\rm nor}(k)=\omega$ [see Fig.\,\ref{fig2}(b)], which arises from the destructive interference between two symmetric channels of the lattice. Notice that cavity $A$ interacts with spins and cavity $B$ both with the form of $x$-$x$ couplings, which formed two symmetric channels (labeled by 1 and 2) as shown in Fig.\,\ref{fig1}(b). The remarkable thing is that this destructive interference leads to zero occupancy of cavity $A$ [see Fig.\,\ref{fig3}(a,c)], which is reminiscent of the dark states known from electromagnetically-induced transparency and coherent population trapping \cite{Arimondo1976,Fleischhauer2005}. To achieve perfect destructive interference, two necessary conditions must be satisfied simultaneously; one is the on-site  resonance condition between spin ensemble and cavity $B$, i.e., $\omega_{B}=\Omega$, and the other is the chiral symmetry of the off-site interaction matrix $\mathcal{M}^{\rm int}_{\rm nor}(k)$, i.e.,  $\{\mathcal{C},\mathcal{M}^{\rm int}_{\rm nor}(k)\}=0$. This determines the present flat band is chiral-symmetry-protected, and thus it is robust against the system parameters $\lambda$ and $\zeta$.


This symmetry-protected flat band persists in the whole normal phase, but immediately disappears once the system enters into superradiant phase with increasing $\lambda$, as shown in Figs.\,\ref{fig2}(d,e). In the superradiant phase, the spin ensemble acquiring macroscopic occupation causes the excitation energy of spins to become $\chi+2\eta$ [see Eq.\,(\ref{matrix2})], which closely depends on the coupling strength $\lambda$. Then the on-site resonant condition of flat band is destroyed, i.e., $\chi+2\eta\ne\omega_{B}$. Moreover, the QPT induces an interaction-dependent potential of spin ensemble, which is transformed into the additional pairings of spins in the momentum space, i.e., $d_{k_A}^{\dag}d_{-k_A}^{\dag}$ and $d_{k_A}d_{-k_A}$ terms. These pairing terms  breaks up the chiral symmetry of off-site matrix $\mathcal{M}^{\rm int}_{\rm sup}(k)$, i.e., $\{\mathcal C,\mathcal{M}^{\rm int}_{\rm sup}(k)\}\ne0$. Both of the above elements prevent the appearance of flat band in the superradiant phase. As shown in Figs.\,\ref{fig2}(c,d), in the critical region between the normal and superradiant phases, the valid flat band appears in the special range of $k$, in that different spreading waves decided by $k$ have different critical points.

Besides the flat band, the present energy spectrum also features asymmetric band structures, which is distinguishable from the case of normal lattice under the rotating wave approximation. Physically this comes from the counter-rotating terms in spin-cavity and cavity-cavity interactions, which induces the non-zero ground-state expectations of the pairing operators between $k$ and $-k$ space, i.e.,  $\langle a_{k_A}b_{-k_A}\rangle_{g}$ ($\langle c_{k_A}d_{-k_A}\rangle_{g}$, $\langle d_{k_A}d_{-k_A}\rangle_{g}$) in the normal (superradiant) phase [see Fig.\,\ref{fig2}(f)].


The above results offer a robust method to identify different phases of quantum lattice system by detecting its energy bands. Normally the flat band can induce periodical population localization, which provides an auxiliary method for identifying QPT via only probing the occupation of sites in a single cell. In Fig.\,{\ref{fig3}}, we numerically calculate the local density of states (LDOS),
$
\rho_n(E)=\sum_{lk}|\langle \chi_n|\phi_{jk}\rangle|^2\delta(E-E_{jk}),
$
where the subscript $n$ differentiates three physical modes, i.e., cavity modes $A$, $B$, and spins. $|\chi_n\rangle$ is the basis state corresponding to occupation of mode $n$. The sum $\sum_{jk}$ runs over various energy bands in the first Brillouin zone, and $E_{jk}$ is the eigenvalue related to eigenstate $|\phi_{jk}\rangle$. Figs.\,{\ref{fig3}}(a) and (b) show the LDOS of cavity modes $A$, $B$ and spin mode with respect to the middle band in the normal phase and superradiant phase, respectively.
In the normal phase, the LDOS of cavity $A$ at the flat band is zero, in other words, there is no particle occupied in site $A$. But for cavity $B$ and spins, the LDOSs have regular Gaussian-like peaks.
In such a regime, both cavity $B$ and spins are localized at the flat band $E=\omega$, while cavity $A$ remains completely dark in that the destructive interference between two channels cancels the net occupations in  cavity mode $A$ [see Fig.\,\ref{fig3}(c)]. Nevertheless, in the superradiant phase, the destructive interference is destroyed, causing the disappearance of flat-band localization. As shown in Figs.\,\ref{fig3}(b) and (d), the cavities $A$, $B$, and spin modes are all occupied.  Therefore, in principle, the different phases of quantum lattice system can be identified by detecting the localization effects of a single cell. 
\section{2D extended Dicke-Hubbard lattice}\label{section3}
\begin{figure*}
	\centerline{\includegraphics[width=15.0cm]{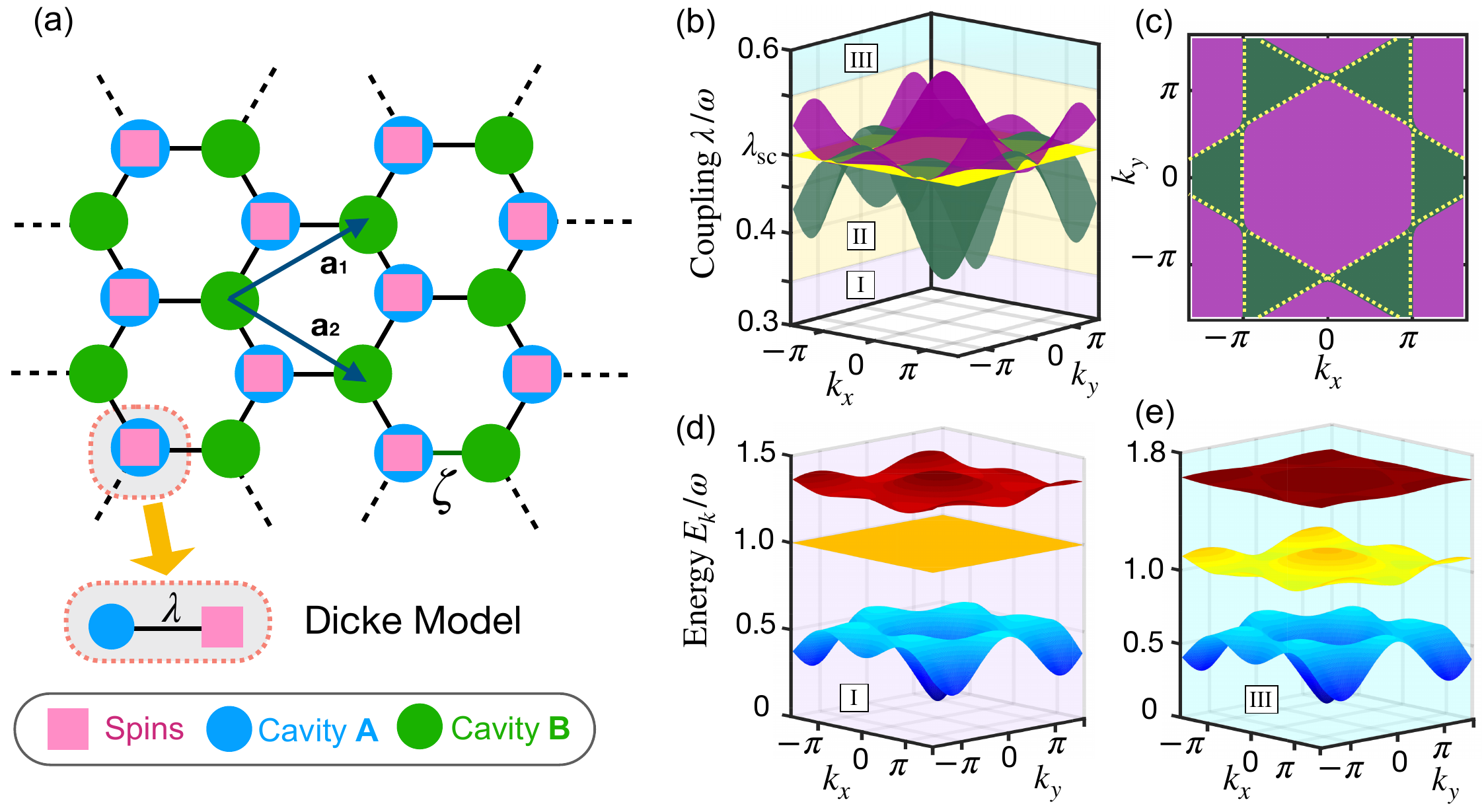}}
	\caption{(a) Schematic illustration of 2D {extended} Dicke-Hubbard lattice with honeycomb structure. (b) The normal phase boundary $E_{\rm nor}^{{\rm 2D}l}(\textbf{k},\lambda)=0$ (bottom branch) and superradiant phase boundary $E_{\rm sup}^{{\rm 2D}l}(\textbf{k},\lambda)=0$ (upper branch) intersect at the plane $\lambda=\lambda_{\rm sc}$ in the 3D critical region labelled by II. The normal and superradiant phases are labeled by I and III, respectively. (c) The top view of (b) projected into the plane $\lambda=\lambda_{\rm sc}$, and the yellow-dotted curves are the analytical intersection curves satisfying $f(k_x,k_y)=0$. (d,e) Energy structure of 2D {extended} Dicke-Hubbard lattice in the normal phase $\lambda=0.34\omega$ and superradiant phase $\lambda=0.58\omega$, respectively.  In the normal phase, the flat band localizes at $E= \omega$. But all energy bands become dispersive in the superradiant phase. Here we consider $\Omega=\omega_A=\omega_B=\omega=1$ and $\zeta=0.12\omega$. }
	\label{fig4}
\end{figure*}
Now we will extend our results to the case of  2D {extended} Dicke-Hubbard lattice, where the established connection between flat band and QPT in the 1D model is still held.  As an example, we consider the 2D  {extended} Dicke-Hubbard lattice with honeycomb structure, as shown in Fig.\,\ref{fig4}(a). We assume the lattice constant $|\textbf{a}|=\sqrt{3}/3$ and thus basis vectors read $\textbf{a}_1=(1,0)$, and $\textbf{a}_2=(1/2,\sqrt{3}/2)$. Different from 1D lattice, here every cavity mode has three nearest neighbors, thus the interaction between two nearest-neighbor cavities reads
\begin{align}
{\mathcal H}^{\rm 2D}_{\rm int}=&-\zeta\sum_{i}\left[\left(a_{A,\textbf{r}_{i}}^{\dagger}+ a_{A,\textbf{r}_i}\right)\left(a_{B,\textbf{r}_i+\textbf{e}_1}^{\dagger}+a_{B,\textbf{r}_i+\textbf{e}_1}\right) \right]\nonumber\\ &-\zeta\sum_{i}\left[\left(a_{A,\textbf{r}_i}^{\dagger}+a_{A,\textbf{r}_i}\right)\left(a_{B,\textbf{r}_i+\textbf{e}_2}^{\dagger}+ a_{B,\textbf{r}_i+\textbf{e}_2}\right)\right]
\nonumber\\&-\zeta\sum_{i}\left[\left(a_{A,\textbf{r}_i}^{\dagger}+a_{A,\textbf{r}_i}\right)\left(a_{B,\textbf{r}_i+\textbf{e}_3}^{\dagger}+a_{B,\textbf{r}_i+\textbf{e}_3}\right)\right],
\end{align}
with
$
\textbf{e}_1=\left(0,{\sqrt{3}}/{3}\right),
\textbf{e}_{2}=\left(-{{1}}/{2}, -{\sqrt{3}}/{6}\right),
\textbf{e}_{3}=\left({{1}}/{2}, -{\sqrt{3}}/{6}\right).
$
Here the sum $\sum_{i}$ runs over all unit cells and $a_{A,\textbf{r}_i}$ ($a_{B,\textbf{r}_i}$) is the annihilation operator for the cavity mode $A$ (cavity mode $B$) and $\textbf {r}_i$ is the position vector in the $i$th unit cell. Then the total Hamiltonian is
\begin{align}
{\mathcal H^{\rm 2D}}=&\sum_{n_A}{\mathcal H}_{n_A}^{\rm Dicke}+\sum_{n_B} {\mathcal H}_{n_B}^{\rm Cavity}+{\mathcal H}^{\rm 2D}_{\rm int}.
\end{align}
We apply the same Holstein-Primakoff representation and displacement process as before [see more details in Sec.\,\ref{sec2}], and perform a 2D Fourier transformation
$
a_{A,\textbf{k}}=\frac{1}{\sqrt{\mathcal N}}\sum_{i}a_{A,\textbf{r}_i}e^{-i\textbf{k}\cdot\textbf{r}},
a_{B,\textbf{k}}=\frac{1}{\sqrt{\mathcal N}}\sum_{i}a_{B,\textbf{r}_i}e^{-i\textbf{k}\cdot\textbf{r}}.
$
Then we obtain the $6\times 6$ coefficient matrix with the same form with Eq.\,({\ref{matrix1}}) in the normal phase and Eq.\,({\ref{matrix2}}) in the superradiant phase but with $f$ replaced by
\begin{align}f^{\rm 2D}=-\zeta\left[1+{\rm exp}\left(i\textbf{k}\cdot \textbf{a}_1\right)+{\rm exp}\left(i\textbf{k}\cdot \textbf{a}_2\right)\right],
\end{align}
where $\textbf{k}=(k_x,k_y)$. 
Diagonalizing the coefficient matrix, we can obtain the eigenvalues of system in the normal phase are
\begin{widetext}
	\begin{eqnarray}
E_{\rm nor}^{{\rm 2D}l}(\textbf{k},\lambda)=&\sqrt{\omega^2-2\omega\sqrt{\zeta^2[3+2 {\rm cos}\,(k_x)+4{\rm cos}\,(\frac{k_x}{2}){\rm cos}\,(\frac{\sqrt{3}k_y}{2})]+\lambda^2}},\\
E_{\rm nor}^{{\rm 2D}m}(\textbf{k},\lambda)=&\omega,\;\;\;\;\;\;\;\;\;\;\;\;\;\;\;\;\;\;\;\;\;\;\;\;\;\;\;\;\;\;\;\;\;\;\;\;\;\;\;\;\;\;\;\;\;\;\;\;\;\;\;\;\;\;\;\;\;\;\;\;\;\;\;\;\;\;\;\;\;\;\;\;\;\;\;\;\;\;\;\;\;\;\\
E_{\rm nor}^{{\rm 2D}h}(\textbf{k},\lambda)=&\sqrt{\omega^2+2\omega\sqrt{\zeta^2[3+2 {\rm cos}\,(k_x)+4{\rm cos}\,(\frac{k_x}{2}){\rm cos}\,(\frac{\sqrt{3}k_y}{2})]+\lambda^2}},\label{eigenval2}
\end{eqnarray}
\end{widetext}
where we have taken $\omega_A=\omega_B=\Omega=\omega$. The eigenvalues of system in the superradiant phase are $E_{\rm sup}^{{\rm 2D}l}(\textbf{k},\lambda)$, $E_{\rm sup}^{{\rm 2D}m}(\textbf{k},\lambda)$, and $E_{\rm sup}^{{\rm 2D}h}(\textbf{k},\lambda)$, which have the complicated form so in the following we show its numerircal form.

\begin{figure*}
	\centerline{\includegraphics[width=15.5cm]{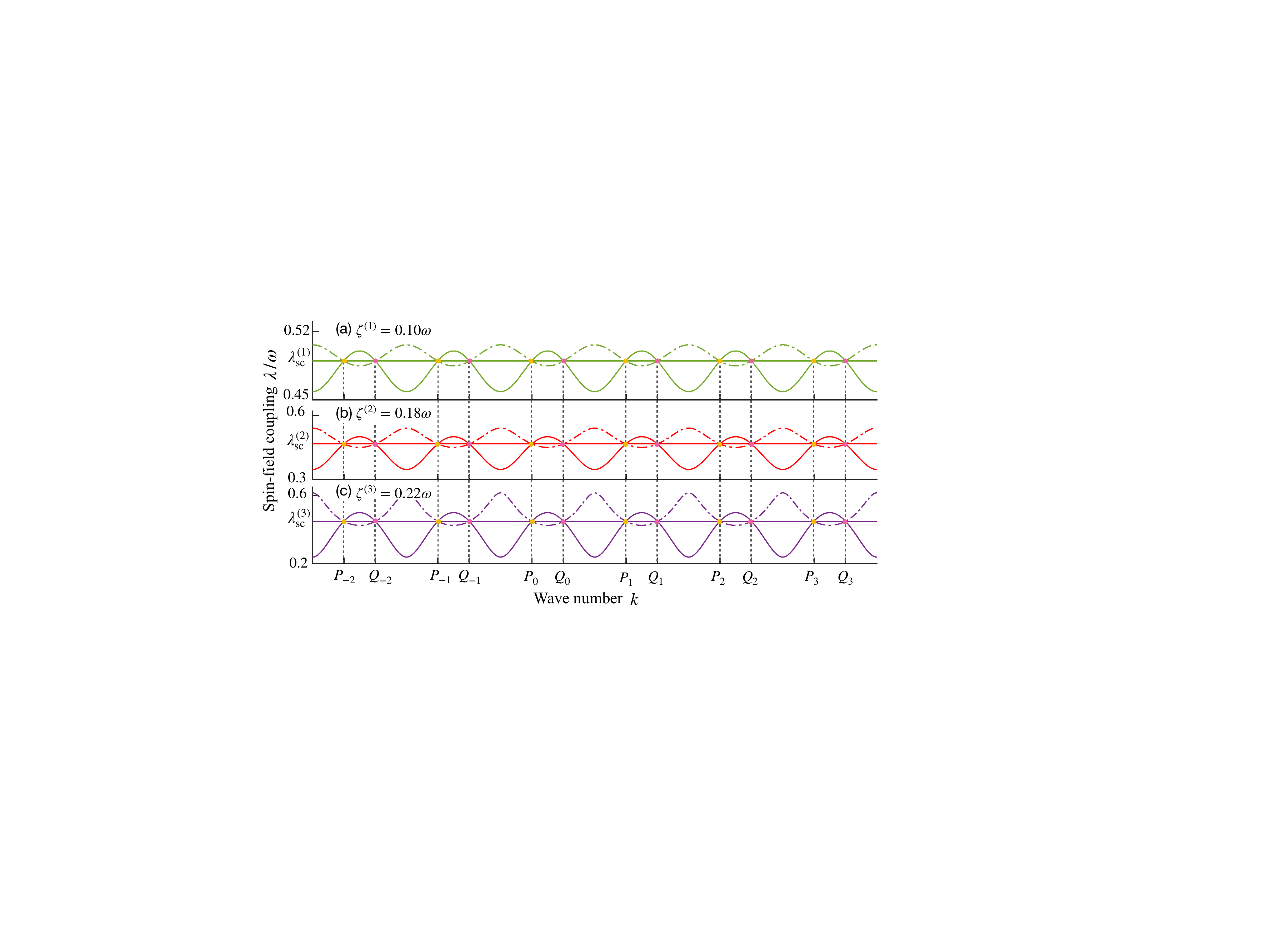}}
	\caption{(a-c) Plots of the normal phase boundary $E_{\rm nor}^{l}(k,\lambda)=0$ (lower solid curves) and superradiant phase boundary $E_{\rm sup}^{l}(k,\lambda)=0$ (upper dash-dotted curves) for $\zeta^{(1)}=0.10\omega, \zeta^{(2)}=0.18\omega, \zeta^{(3)}=0.22\omega$. They intersect at the middle straight lines $\lambda=\lambda^{(i)}_{\rm sc}$, where $\lambda_{\rm sc}^{(i)}=\sqrt{\Omega(\omega_A-4\zeta^{(i)2}/\omega_B)}/2$ ($i=1,2,3$) are the critical points of QPT occurring in a single cell. Here $P_n=2n\pi-4\pi/3,\,Q_n=2n\pi-2\pi/3$ are the $x$-coordinates of the crossing points, and $\Omega=\omega_A=\omega_B=1$ are considered.}
	\label{fig5}
\end{figure*}

As show in Fig. {\ref{fig4}(b)}, the 2D {extended} Dicke-Hubbard lattice has similar QPT property as the case of 1D shown in Fig.\,\ref{fig2}(a). Specifically, there are three parameter regions, i.e., the normal phase, critical phase, and superradiant phase, with increasing $\lambda$ in the 3D parameter space in terms of $\lambda$, $k_x$, and $k_y$.  Now the boundaries of  normal and superradiant phases, i.e., $E_{\rm nor}^{{\rm 2D}l}(\textbf{k},\lambda)=0$ and $E_{\rm sup}^{{\rm 2D}l}(\textbf{k},\lambda)=0$, overlap at the plane $\lambda=\lambda_{\rm sc}$. The overlaped curves satisfy $ f(k_x,k_y)={\rm cos}\,(k_x)+2{\rm cos}\,({k_x}/{2}){\rm cos}\,({\sqrt{3}k_y}/{2})+1=0,$ which is obtained by plugging $\lambda=\lambda_{\rm sc}$ into the surface diagram $E_{\rm nor}^{{\rm 2D}l}(\textbf{k},\lambda)=0$. This result is clearly demonstrated in Fig.\,\ref{fig4}(c), where the analytical overlaped curves [yellow-dotted curves corresponding to $f(k_x,k_y)=0$] are good consistent with the numerical results. 

To show the connection between flat band and QPT in the present 2D model, in Figs.\,{\ref{fig4}}(d,e), we plot band structure of system in normal and superradiant phases, respectively. It is clearly shown that energy spectra own three doubly-degenerated bands, and there is a flat band exhibiting at $E=\omega$ in the normal phase. Physically, the chiral symmetry of system and initial resonant condition still maintain in the 2D {extended} Dicke-Hubbard lattice, which allows the flat band persists in the normal phase. But in the superradiant phase, the macroscopic excitations introduce interaction-dependent excitation energy and additional potential of spin ensemble, breaking up the on-site resonate condition and off-site chiral symmetry of lattice, respectively, which makes all energy bands dispersive. This result shows a well agreement with the 1D {extended} Dicke-Hubbard lattice. 
\section{discussions and conclusions}
Regarding experimental implementations, while we have considered here a hybrid superconducting circuit with the ensembles of spins in diamond crystal coupled to the transmission line resonators (forming the Dicke Model)\,\cite{Kubo2010,Schuster2010,Amsuss2011,Ranjan2013,Astner2017,Grezes2014}, our proposal is not limited to this particular architecture, and could be implemented or adapted in a variety of platforms, e.g., atomic\,\cite{Weiss2015,Verdu2009}, molecular\,\cite{Rabl2006} and ferromagnetic\,\cite{Huebl2013,Tabuchi2015,You2018,Li2019} systems coupled to superconducting cavities. For our specific design, considering an ensemble with $N\sim 10^{12}$ spins and the single spin coupling $\lambda_0\sim10{\rm Hz}$, an enhanced collective coupling $\lambda\approx \sqrt{N}\lambda_0\sim 10 {\rm MHz}$ \,\cite{Kubo2010,Schuster2010,Amsuss2011,Ranjan2013,Astner2017,Grezes2014} allows our model to reach ultra-strong coupling regime, which demonstrates that the critical coupling of QPT can be readily realized with stat-of-the-art technology. To experimentally detect band structure, one of the most generally used techniques is photoluminescence \cite{Jacqmin2014,Baboux2016}. The emission of the sample can be collected through a high numerical aperture objective, dispersed in a spectrometer and detected by a CCD camera on which the energy structure of either the real or the momentum space can be directly imaged. Alternatively, as reported in Ref.\,\cite{Bellec2019}, a loop antenna mounted on a scanning system connected to a vectorial network analyzer, can be used to collect the signal both spectrally and spatially resolved and further allows to obtain the local density of states.
 
We have investigated the quantum critical and energy band properties of an extended Dicke-Hubbard lattice and established the connection between the flat band and the superradiant phase transition.
Comparing to the single Dicke model, the extended Dicke-Hubbard lattice features rich equilibrium dynamics dependent on the wave number $k$, including the periodical boundaries of phases, and a critical region between normal and superradiant phases. More importantly, we found the symmetry-protected flat band and the population localization are observed exclusively in the normal phase of the system, which offers a robust method to detect the phase of the system, as well as manipulate the flat band of the lattice with QPT.
Our work would inspire further exploration regarding the interactions between quantum properties associating with QPT (e.g., quantum entanglement, quantum chaos\,\cite{Emary2003,Lambert2004}), the dynamical phase transitions\,\cite{Baumann2010,Nagy2010,Klinder2015}  and energy band theory in the lattice systems. 
\begin{acknowledgments}
We thank T. Shi, Q. Bin, L. L. Wan, Z. H. Li, and Z. Gao for helpful discussions. This work is supported by the National Science Foundation of China (Grant Nos. 11822502, 11974125, and 11875029), the National Key Research and Development Program of China grant 2016YFA0301203, and the Fundamental Research Funds for the Central Universities grant No. 2019kfyXMBZ054.
\end{acknowledgments}

\appendix

	\section{Derivation of the intersections in the critical region}\label{section2}
	In this section, we first derive the critical point of QPT in the single unit cell. The Hamiltonian in the $i$th unit cell is 
	\begin{align}
	{\mathcal H_i}=&\omega_Aa_{i_A}^{\dag}a_{i_A}+\Omega J_{i_A}^{z}+\frac{\lambda}{\sqrt{N}}\left(a_{i_A}^{\dag}+a_{i_A}\right)\left(J_{i_A}^{+}+J_{i_A}^{-}\right)\nonumber\\&+\omega_B a_{i_B}^{\dag}a_{i_B}-\zeta\left(a_{i_A}^{\dag}+a_{i_A}\right)\left(a_{i_B}^{\dag}+a_{i_B}\right),
	\end{align}
where the third term describe $N$ spins collectively interact with single cavity mode with the coupling $\lambda$ and the last term is the interaction between nearest-neighbor cavities.
Using the method proposed in Ref.\,\cite{Emary2003}, in a single unit cell we can obtain the critical point as
\begin{align}
\lambda_{\rm sc}=\frac{\sqrt{\Omega(\omega_A-4\zeta^2/\omega_B)}}{2}.
\label{EQ:cconecell}
\end{align}
Different from the normal Dicke model, the interaction between nearest-neighbor cavities introduces a shift towards smaller spin-field couplings for the critical point in a single unit cell. 

Based on the above critical point of a single cell, let us derive the coordinates of crossing points in the critical region of the lattice system, i.e., ($P_n$, $\lambda_{\rm sc}$) and ($Q_n$, $\lambda_{\rm sc}$) shown in Fig.\,{\ref{fig5}}. Physically, when the periodical boundaries of normal and superradiant phases, i.e., $E_{\rm nor}^{l}(k,\lambda)=0$ and $E_{\rm sup}^{l}(k,\lambda)=0$, intersect at the special values of $k$, our lattice system undergoes a phase transition from normal phase directly to the superradiant phase with increasing $\lambda$, which is same as the case of a single cell. Therefore, the $\lambda$-coordinates of the crossing points should be $\lambda_{\rm sc}$, and in other words, the curves $E_{\rm nor}^{l}(k,\lambda)=0$, $E_{\rm sup}^{l}(k,\lambda)=0$ and the straight line $\lambda=\lambda_{\rm sc}$ should touch at the same point, which is clearly demonstrated in Fig.\,\ref{fig5}. Therefore, plugging $\lambda=\lambda_{\rm sc}$ into the curves $E_{\rm nor}^{l}(k,\lambda)=0$ [Eq. (\ref{eigenval1})] or $E_{\rm sup}^{l}(k,\lambda)=0$ [Eq.(\ref{eigenval2})], we can obtain the $k$-coordinates of the crossing points satisfying ${\rm cos\,}k=-1/2$. Lastly, in the critical region of the lattice system, the normal and superradiant phase boundaries intersect at ($P_n$, $\lambda_{\rm sc}$) and ($Q_n$, $\lambda_{\rm sc}$) with $P_n=2n\pi-4\pi/3$ and $Q_n=2n\pi-2\pi/3$ ($n=0,\pm1,\pm2...$).


\begin{thebibliography}{100}

\bibitem{Macieszczak2016} K. Macieszczak, M. Guta, I. Lesanovsky, and J. P. Garrahan, Dynamical phase transitions as a resource for quantum enhanced metrology, Phys. Rev. A \textbf{93}, 022103 (2016).

\bibitem{Lorenzo2017} S. F.-Lorenzo and D. Porras, Quantum sensing close to a dissipative phase transition: Symmetry breaking and criticality as metrological resources,
Phys. Rev. A \textbf{96}, 013817 (2017).

\bibitem{Farbe} L. Garbe, M. Bina, A. Keller, M. G. A. Paris, and S. Felicetti, Critical Quantum metrology with a finite-component quantum phase transition, arXiv:1910.00604 [quant-ph].

\bibitem{Goldstein2010} M. Goldstein, R. Berkovits, and Y. Gefen, Population Switching and Charge Sensing in Quantum Dots: A Case for a Quantum Phase Transition,
Phys. Rev. Lett. \textbf{104}, 226805 (2010).

\bibitem{Zhang2018} T. Zhang, G.-Q. Liu, W.-H. Leong, C.-F. Liu, M.-H. Kwok, T. Ngai, R.-B. Liu and Q. Li, Hybrid nanodiamond quantum sensors enabled by volume phase transitions of hydrogels, Nat. Commun. \textbf{9}, 3188 (2018).

\bibitem{Raghunandan2018} M. Raghunandan, J. Wrachtrup, and H. Weimer, High-Density Quantum Sensing with Dissipative First Order Transitions, Phys. Rev. Lett. \textbf{120}, 150501 (2018).
{\color{black}\bibitem{You2017} X.-Y. Luo, Y.-Q. Zou, L.-N. Wu, Q. Liu, M.-F. Han, M. K. Tey, and L. You, Deterministic entanglement generation from driving through quantum phase transitions, Science \textbf{355}, 620 (2017).}

\bibitem{Huang2018} J. Huang, M. Zhuang, and C. Lee, Non-Gaussian precision metrology via driving through quantum phase transitions, Phys. Rev. A \textbf{97}, 032116 (2018).	

\bibitem{Sachdev} S. Sachdev, \emph{Quantum Phase Transitions} (Cambridge University Press, Cambridge, 1999).











\bibitem{Lieb1989} E. H. Lieb, Two theorems on the Hubbard model, Phys. Rev. Lett. \textbf{62}, 1201 (1989).

\bibitem{Vidal1998} J. Vidal, R. Mosseri, and B. Doucot, Aharonov-Bohm Cages in Two-Dimensional Structures, Phys. Rev. Lett. \textbf{81}, 5888 (1998).

\bibitem{Bergman2008} D. L. Bergman, C. Wu, and L. Balents, Band touching from real-space topology in frustrated hopping models, Phys. Rev. B \textbf{78}, 125104 (2008).

\bibitem{Ge2015} L. Ge, Parity-time symmetry in a flat-band system, Phys. Rev. A \textbf{92}, 052103 (2015).

\bibitem{Qiu2016} W.-X. Qiu, S. Li, J.-H. Gao, Y. Zhou, and F.-C. Zhang, Phys. Rev. B \textbf{94}, 241409(R) (2016). 

\bibitem{Ozawa2017} H. Ozawa, S. Taie, T. Ichinose, and Y. Takahashi, Interaction-Driven Shift and Distortion of a Flat Band in an Optical Lieb Lattice, Phys. Rev. Lett. \textbf{118}, 175301 (2017).

\bibitem{Hamid2017} H. Ramezani, Non-Hermiticity-induced flat band, Phys. Rev. A \textbf{96}, 011802(R) (2017).


\bibitem{Tang2011} E. Tang, J.-W. Mei, and X.-G. Wen, High-Temperature Fractional Quantum Hall States, Phys. Rev. Lett. \textbf{106}, 236802 (2011).

\bibitem{Sun2011} K. Sun, Z. Gu, H. Katsura, and S. Das Sarma, Nearly Flatbands with Nontrivial Topology, Phys. Rev. Lett. \textbf{106}, 236803 (2011).

\bibitem{Neupert2011} T. Neupert, L. Santos, C. Chamon, and C. Mudry, Fractional Quantum Hall States at Zero Magnetic Field, Phys. Rev. Lett. \textbf{106}, 236804 (2011).

\bibitem{Sheng2011} D. N. Sheng, Z.-C. Gu, K. Sun, and L. Sheng, Fractional quantum Hall effect in the absence of Landau levels, Nat. Commun. \textbf{2}, 389 (2011).

\bibitem{Regnault2011} N. Regnault and B. A. Bernevig, Fractional Chern Insulator, Phys. Rev. X \textbf{1}, 021014 (2011).

\bibitem{Biondi2015} M. Biondi, E. P. L. van Nieuwenburg, G. Blatter, S. D. Huber, and S. Schmidt, Incompressible Polaritons in a Flat Band, Phys. Rev. Lett. \textbf{115}, 143601 (2015).

\bibitem{silva2014}D. Guzm\'{a}n-Silva {\it et al.}, Experimental observation of bulk and edge transport in photonic Lieb lattices, New J. Phys. \textbf{16}, 063061 (2014).

\bibitem{Jacqmin2014} T. Jacqmin, I. Carusotto, I. Sagnes, M. Abbarchi, D. D. Solnyshkov, G. Malpuech, E. Galopin, A. Lema\^{\i}tre, J. Bloch, and A. Amo, Direct Observation of Dirac Cones and a Flatband in a Honeycomb Lattice for Polaritons, Phys. Rev. Lett. \textbf{112}, 116402 (2014).

\bibitem{Baboux2016} F. Baboux, L. Ge, T. Jacqmin, M. Biondi, E. Galopin, A. Lema\^{\i}tre, L. Le Gratiet, I. Sagnes, S. Schmidt, H. E. T\"{u}reci, A. Amo, and J. Bloch, Bosonic Condensation and Disorder-Induced Localization in a Flat Band, Phys. Rev. Lett. \textbf{116}, 066402 (2016).

\bibitem{Vicencio2015} R. A. Vicencio {\it et al.}, Observation of Localized States in Lieb Photonic Lattices, Phys. Rev. Lett. \textbf{114}, 245503 (2015).

\bibitem{Mukheriee2015} S. Mukherjee, A. Spracklen, D. Choudhury, N. Goldman, P. \"{O}hberg, E. Andersson, and R. R. Thomson, Observation of a Localized Flat-Band State in a Photonic Lieb Lattice, Phys. Rev. Lett. \textbf{114}, 245504 (2015).

\bibitem{alex19} T. Biesenthal, M. Kremer, M. Heinrich, and A. Szameit, Experimental Realization of PT-Symmetric Flat Bands, Phys. Rev. Lett. \textbf{123}, 183601 (2019).


\bibitem{Li2008}J. Li, T. P. White, L. O'Faolain, A. Gomez-Iglesias, and T. F. Krauss, Systematic design of flat band slow light in photonic crystal waveguides, Opt. Express, \textbf{16}, (9) 6227 (2008).

\bibitem{Taie2015} S. Taie, H. Ozawa, T. Ichinose, T. Nishio, S. Nakajima and Y. Takahashi, Coherent driving and freezing of bosonic matter wave in an optical Lieb lattice, Sci. Adv. \textbf{1}, 1500854 (2015).

\bibitem{Dicke1954} R. H. Dicke, Coherence in spontaneous radiation processes, Phys. Rev. \textbf{93}, 99 (1954).
    
\bibitem{Bensky2011} G. Bensky, R. Ams\"{u}ss, J. Majer, D. Petrosyan, J. Schmiedmayer, and G. Kurizki, Controlling quantum information processing in hybrid systems on chips, Quantum Inf. Process. \textbf{10}, 1037 (2011).

\bibitem{Houck2012} A. A. Houck, H. E. T\:{u}reci, and J. Koch, On-chip quantum simulation with superconducting circuits, Nature Physics \textbf{8}, 292–299 (2012). 

\bibitem{Underwood2012} D. L. Underwood, W. E. Shanks, J. Koch, and A. A. Houck, Low-disorder microwave cavity lattices for quantum simulation with photons, Phys. Rev. A \textbf{86}, 023837 (2012).

\bibitem{Ze-Liang} Z.-L. Xiang, S. Ashhab, J. Q. You, and F. Nori, Hybrid quantum circuits: superconducting circuits interacting with other quantum systems, Rev. Mod. Phys. \textbf{85}, 623--653 (2013).

\bibitem{xinyou2013} X.-Y. L\"{u}, Z.-L. Xiang, W. Cui, J. Q. You, and F. Nori, Quantum memory using a hybrid circuit with flux qubits and nitrogen-vacancy centers, Phys. Rev. A \textbf{88}, 012329 (2013). 

\bibitem{Zou2014}L. J. Zou, D. Marcos, S. Diehl, S. Putz, J. Schmiedmayer, J. Majer, and P. Rabl, Implementation of the Dicke Lattice Model in Hybrid Quantum System Arrays, Phys. Rev. Lett. \textbf{113}, 023603 (2014).

\bibitem{Li2015} P.-B. Li, Y.-C. Liu, S.-Y. Gao, Z.-L. Xiang, P. Rabl, Y.-F. Xiao, and F.-L. Li, Hybrid Quantum Device Based on NV Centers in Diamond Nanomechanical Resonators Plus Superconducting Waveguide Cavities, Phys. Rev. Applied \textbf{4}, 044003 (2015).

\bibitem{Hepp1973} K. Hepp and E. H. Lieb, Ann. Phys. (N.Y.) \textbf{76}, 360 (1973).

\bibitem{Wang1973} Y. K. Wang and F. T. Hioe, Phase Transition in the Dicke Model of Superradiance, Phys. Rev. A \textbf{7}, 831 (1973).

\bibitem{Emary2003} C. Emary and T. Brandes, Quantum chaos triggered by precursors of a quantum phase transition: the Dicke model, Phys. Rev. Lett. \textbf{90}, 044101 (2003); Chaos and the quantum phase transition in the Dicke model, Phys. Rev. E. \textbf{67}, 066203 (2003).

\bibitem{Lambert2004} N. Lambert, C. Emary, and T. Brandes, Entanglement and the Phase Transition in Single-Mode Superradiance, { Phys. Rev. Lett.} {\bf 92}, 073602 (2004).

\bibitem{Baumann2010} K. Baumann, C. Guerlin, F. Brennecke, and T. Esslinger, Dicke quantum phase transition with a superfluid gas in an optical cavity, { Nature} {\bf 464}, 1301 (2010).

\bibitem{Nagy2010} D. Nagy, G. K\'{o}nya, G. Szirmai, and P. Domokos, Dicke-Model Phase Transition in the Quantum Motion of a Bose-Einstein Condensate in an Optical Cavity, Phys. Rev. Lett. \textbf{104}, 130401 (2010).

\bibitem{Nataf2010} P. Nataf and C. Ciuti, No-go theorem for superradiant quantum phase transitions in cavity QED and counter-example in circuit QED, Nat. Commun. \textbf{1}, 1 (2010).

\bibitem{Klinder2015} J. Klinder, H. Keler, M. Wolke, L. Mathey, and A. Hemmerich, Dynamical phase transition in the open Dicke model, PNAS \textbf{112}, 3290 (2015).

\bibitem{Hanpu2016} C. Zhu, L. Dong, and H. Pu, Effects of spin-orbit coupling on Jaynes-Cummings and Tavis-Cummings models, Phys. Rev. A \textbf{94}, 053621 (2016).

\bibitem{Liu2018} M. Liu, S. Chesi, Z.-J. Ying, X. Chen, H.-G. Luo, and H.-Q. Lin, Universal Scaling and Critical Exponents of the Anisotropic Quantum Rabi Model, Phys. Rev. Lett. \textbf{119}, 220601 (2017).

\bibitem{Xin-You2018} X.-Y. L\"{u}, L.-L. Zheng, G.-L. Zhu, and Y. Wu, Single-Photon-Triggered Quantum Phase Transition, Phys. Rev. Applied {\bf 9}, 064006 (2018).

\bibitem{Guilei2019} G.-L. Zhu, X.-Y. L\"{u}, L.-L. Zheng, Z.-M. Zhan, F. Nori, and Y. Wu, Single-photon-triggered quantum chaos,
Phys. Rev. A \textbf{100}, 023825 (2019).

\bibitem{Hanpu2019} Y. Xu and H. Pu, Emergent Universality in a Quantum Tricritical Dicke Model, Phys. Rev. Lett. \textbf{122}, 193201 (2019).

\bibitem{Emary2002}C. Emary and R. F. Bishop, Exact isolated solutions for the two-photon Rabi Hamiltonian, J. Phys. A: Math. Gen. \textbf {35}, 8231 (2002).

\bibitem{Hayn2011} M. Hayn, C. Emary, and T. Brandes, Phase transitions and dark-state physics in two-color superradiance, Phys. Rev. A \textbf{84}, 053856 (2011); Superradiant phase transition in a model of three-level-$\Gamma$ systems interacting with two bosonic modes, Phys. Rev. A \textbf{86}, 063822 (2012).

\bibitem{Baksic2014}A. Baksic and C. Ciuti, Controlling Discrete and Continuous Symmetries in Superradiant Phase Transitions with Circuit QED Systems, Phys. Rev. Lett. \textbf{112}, 173601
(2014).

\bibitem{Soriente2018} M. Soriente, T. Donner, R. Chitra, and O. Zilberberg, Dissipation-induced anomalous multicritical phenomena, Phys. Rev. Lett. \textbf{120}, 183603 (2018).

\bibitem{Yang2020} C. J. Zhu, L. L. Ping, Y. P. Yang, and G. S. Agarwal, Squeezed light induced symmetry breaking superradiant phase transition, Phys. Rev. Lett. \textbf{124}, 073602 (2020).

\bibitem{Arimondo1976} E. Arimondo and G. Orriols, Nonabsorbing atomic coherences by coherent two-photon transitions in a three-level optical pumping, Lett. Nuovo Cimento \textbf{17}, 333
(1976).


\bibitem{Fleischhauer2005} M. Fleischhauer, A. Imamoglu, and J. P. Marangos, Electromagnetically induced transparency: Optics in coherent media, Rev. Mod. Phys. \textbf{77}, 633 (2005).




\bibitem{Kubo2010} Y. Kubo, F. R. Ong, P. Bertet, D. Vion, V. Jacques, D. Zheng, A. Dréau, J.-F. Roch, A. Auffeves, F. Jelezko, J. Wrachtrup, M. F. Barthe, P. Bergonzo, and D. Esteve, Strong Coupling of a Spin Ensemble to a Superconducting Resonator, Phys. Rev. Lett. \textbf{105}, 140502 (2010).

\bibitem{Schuster2010} D. I. Schuster, A. P. Sears, E. Ginossar, L. DiCarlo, L. Frunzio, J. J. L. Morton, H. Wu, G. A. D. Briggs, B. B. Buckley, D. D. Awschalom, and R. J. Schoelkopf, High-Cooperativity Coupling of Electron-Spin Ensembles to Superconducting Cavities, Phys. Rev. Lett. \textbf{105}, 140501 (2010).



\bibitem{Amsuss2011} R. Amsüss, Ch. Koller, T. Nöbauer, S. Putz, S. Rotter, K. Sandner, S. Schneider, M. Schramböck, G. Steinhauser, H. Ritsch, J. Schmiedmayer, and J. Majer, Cavity QED with Magnetically Coupled Collective Spin States, Phys. Rev. Lett. \textbf{107}, 060502 (2011).

\bibitem{Ranjan2013} V. Ranjan, G. de Lange, R. Schutjens, T. Debelhoir, J. P. Groen, D. Szombati, D. J. Thoen, T. M. Klapwijk, R. Hanson, and L. DiCarlo, Probing Dynamics of an Electron-Spin Ensemble via a Superconducting Resonator, Phys. Rev. Lett. \textbf{110}, 067004 (2013).

\bibitem{Grezes2014}C. Grezes, B. Julsgaard, Y. Kubo, M. Stern, T. Umeda, J. Isoya, H. Sumiya, H. Abe, S. Onoda, T. Ohshima, V. Jacques, J. Esteve, D. Vion, D. Esteve, K. Mølmer, and P. Bertet, Multimode Storage and Retrieval of Microwave Fields in a Spin Ensemble,
Phys. Rev. X \textbf{4}, 021049 (2014).

\bibitem{Astner2017} T. Astner, S. Nevlacsil, N. Peterschofsky, A. Angerer, S. Rotter, S. Putz, J. Schmiedmayer, and J. Majer, Coherent Coupling of Remote Spin Ensembles via a Cavity Bus,
Phys. Rev. Lett. \textbf{118}, 140502 (2017).

\bibitem{Verdu2009} J. Verdú, H. Zoubi, C. Koller, J. Majer, H. Ritsch, and J.
Schmiedmayer, Strong Magnetic Coupling of an Ultracold Gas to a Superconducting Waveguide Cavity, Phys. Rev. Lett. \textbf{103}, 043603 (2009).


\bibitem{Weiss2015} P. Weiss, M. Knufinke, S. Bernon, D. Bothner, L. Sárkány, C. Zimmermann, R. Kleiner, D. Koelle, J. Fortágh, and H.
Hattermann, Sensitivity of Ultracold Atoms to Quantized Flux in a Superconducting Ring, Phys. Rev. Lett. \textbf{114}, 113003 (2015).

\bibitem{Rabl2006} P. Rabl, D. DeMille, J. M. Doyle, M. D. Lukin, R. J. Schoelkopf, and P. Zoller, Hybrid Quantum Processors: Molecular Ensembles as Quantum Memory for Solid State Circuits, Phys. Rev. Lett. \textbf{97}, 033003 (2006).

\bibitem{Huebl2013} H. Huebl, C. W. Zollitsch, J. Lotze, F. Hocke, M. Greifenstein, A. Marx, R. Gross, and S. T. B. Goennenwein, High
Cooperativity in Coupled Microwave Resonator Ferrimagnetic Insulator Hybrids, Phys. Rev. Lett. \textbf{111}, 127003 (2013).

\bibitem{Tabuchi2015} Y. Tabuchi, S. Ishino, A. Noguchi, T. Ishikawa, R.
Yamazaki, K. Usami, and Y. Nakamura, Coherent coupling between a ferromagnetic magnon and a superconducting qubit, Science \textbf{349}, 405 (2015).

\bibitem{You2018} Y.-P. Wang, G.-Q Zhang, D. Zhang, T.-F. Li, C.-M. Hu, and J. Q. You, Bistability of Cavity Magnon Polaritons, Phys. Rev. Lett. \textbf{120}, 057202 (2018).

\bibitem{Li2019} Y. Li, T. Polakovic, Y.-L. Wang, J. Xu, S. Lendinez, Z. Zhang, J. Ding, T. Khaire, H. Saglam, R. Divan, J. Pearson, W. K. Kwok, Z. Xiao, V. Novosad, A. Hoffmann, and W. Zhang,
Strong Coupling between Magnons and Microwave Photons in On-Chip Ferromagnet-Superconductor Thin-Film Devices, Phys. Rev. Lett. \textbf{123}, 107701 (2019).

\bibitem{Bellec2019} M. Bellec, C. Poli, U. Kuhl, F. Mortessagne, and H. Schomerus, Observation of supersymmetric pseudo-Landau levels in strained microwave graphene, arXiv:2001.10287.

\end{thebibliography}
\end{document}